\documentstyle[times, natbib, natbibspacing, graphicx, amsmath, amssymb]{jaa}

\def\bibfont{\footnotesize}

\def\memras{\ref@jnl{MmRAS}}            
             
\def\pra{\ref@jnl{Phys.~Rev.~A}}        
\def\prb{\ref@jnl{Phys.~Rev.~B}}        
\def\prc{\ref@jnl{Phys.~Rev.~C}}        
\def\prd{\ref@jnl{Phys.~Rev.~D}}        
\def\pre{\ref@jnl{Phys.~Rev.~E}}        
\def\prl{\ref@jnl{Phys.~Rev.~Lett.}}

\def\qjras{\ref@jnl{QJRAS}}             
\def\skytel{\ref@jnl{S\&T}}             
\def\solphys{\ref@jnl{Sol.~Phys.}}      
\def\sovast{\ref@jnl{Soviet~Ast.}}      
\def\ssr{\ref@jnl{Space~Sci.~Rev.}}     
\def\zap{\ref@jnl{ZAp}}                 
              
\def\iaucirc{\ref@jnl{IAU~Circ.}}
\def\aplett{\ref@jnl{Astrophys.~Lett.}}
\def\apspr{\ref@jnl{Astrophys.~Space~Phys.~Res.}}
\def\bain{\ref@jnl{Bull.~Astron.~Inst.~Netherlands}}
\def\fcp{\ref@jnl{Fund.~Cosmic~Phys.}}
\def\gca{\ref@jnl{Geochim.~Cosmochim.~Acta}}
\def\grl{\ref@jnl{Geophys.~Res.~Lett.}}
\def\jcp{\ref@jnl{J.~Chem.~Phys.}}      
\def\jgr{\ref@jnl{J.~Geophys.~Res.}}    
\def\jqsrt{\ref@jnl{J.~Quant.~Spec.~Radiat.~Transf.}}
\def\memsai{\ref@jnl{Mem.~Soc.~Astron.~Italiana}}
\def\nphysa{\ref@jnl{Nucl.~Phys.~A}}

\def\physscr{\ref@jnl{Phys.~Scr}}
\def\planss{\ref@jnl{Planet.~Space~Sci.}}
\def\procspie{\ref@jnl{Proc.~SPIE}}

\begin{document}
\title[ ]{Relics as probes of  galaxy cluster mergers} 
\author[R.~J. van Weeren et al.]%
       {R.~J. van Weeren$^{1}$\thanks{e-mail:rvweeren@strw.leidenuniv.nl}, M. Br\"uggen$^{2}$  \\
     $^{1}$Leiden Observatory, Leiden University, P.O. Box 9513, NL-2300 RA Leiden, The Netherlands\\
       $^{2}$Jacobs University Bremen, P.O. Box 750561, 28725 Bremen, Germany\\
       \newauthor  H.~J.~A. R\"ottgering$^{1}$, M. Hoeft$^{3}$ \\ 
       $^{3}$Th\"uringer Landessternwarte Tautenburg, Sternwarte 5, 07778, Tautenburg, Germany\\
       }

\maketitle
\label{firstpage}
\begin{abstract}
Galaxy clusters grow by mergers with other clusters and galaxy groups. These mergers create shocks within the intracluster medium (ICM). It is proposed that within the shocks particles can be accelerated to extreme energies. In the presence of a magnetic field these particles should then form large regions emitting synchrotron radiation, creating so-called radio relics. 
An example of a cluster with relics is CIZA~J2242.8+5301. Here we present hydrodynamical simulations of idealized binary cluster collisions with the aim of  constraining the merger scenario for this cluster. We conclude that by using the location, size and width of double radio relics we can set constraints on the mass ratios, impact parameters, timescales, and viewing geometries of binary cluster merger events. 
\end{abstract}

\section{Introduction}
\label{sec:intro}
Radio relics are elongated filamentary sources found in massive merging galaxy clusters. These radio sources indicate the presence of magnetic fields and particle acceleration within the ICM \citep[e.g.,][]{1977ApJ...212....1J, 2004IJMPD..13.1549G}. Since all giant radio relics are found in merging clusters \citep[e.g.,][]{2004ApJ...605..695G, 2007A&A...467...37B, 2009A&A...507..661B, 2010ApJ...721L..82C}, it has been proposed that a small fraction of the energy released during a cluster merger event is channeled into (re)acceleration of particles.
The currently favored scenario for the origin of giant radio relics is that they trace shock waves within the ICM in which particles are (re)accelerated by the diffusive shock acceleration mechanism \citep[e.g.,][]{1983RPPh...46..973D, 1987PhR...154....1B, 1991SSRv...58..259J, 2001RPPh...64..429M}. An alternative scenario has been proposed by \cite{2010arXiv1011.0729K} based on a single secondary cosmic ray electron model, where the time evolution of both magnetic fields and cosmic ray distribution are taken into account to explain giant relics (and radio halos). 

Particularly interesting are the so-called double-relics, where two relics are diametrically located on both sides of the cluster center 
\citep[e.g., ][]{2009A&A...494..429B, 2009A&A...506.1083V, 2006Sci...314..791B, 2010Sci...330..347V,2011ApJ...727L..25B,2011A&A...528A..38V}. These relics are thought to trace outward moving shock waves from a binary cluster merger event, which develop after core passage of the two subclusters. The idea is that double relics can be used to put constraints on the merger timescale, mass ratio, impact parameter and viewing angle.

\section{Simulating binary cluster mergers and double radio relics}
\label{sec:binary}
\begin{figure*}
\includegraphics[angle =90, trim =0cm 0cm 0cm 0cm,width=0.49\textwidth, clip=true]{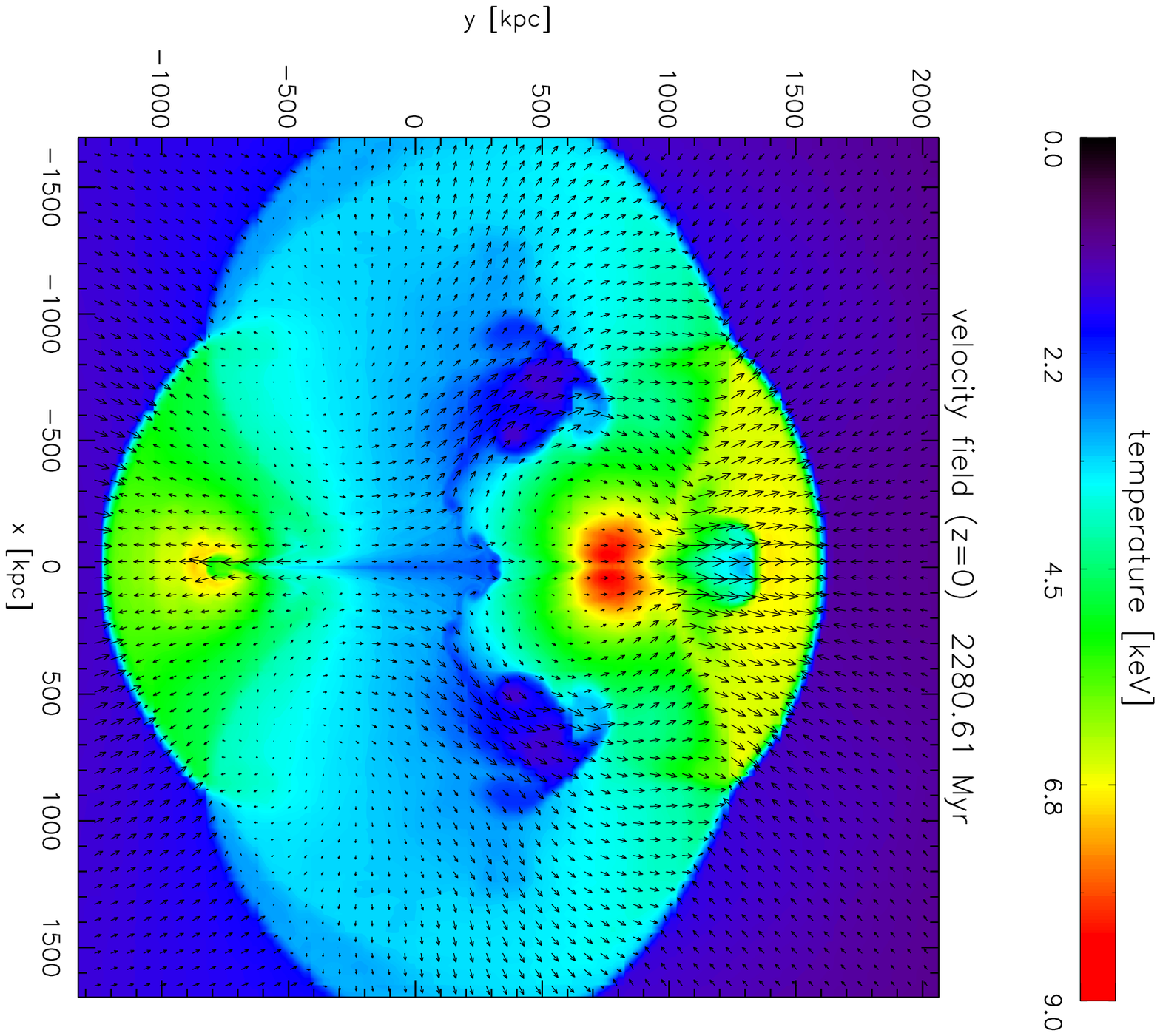}
\includegraphics[angle =90, trim =0cm 0cm 0cm 0cm,width=0.49\textwidth, clip=true]{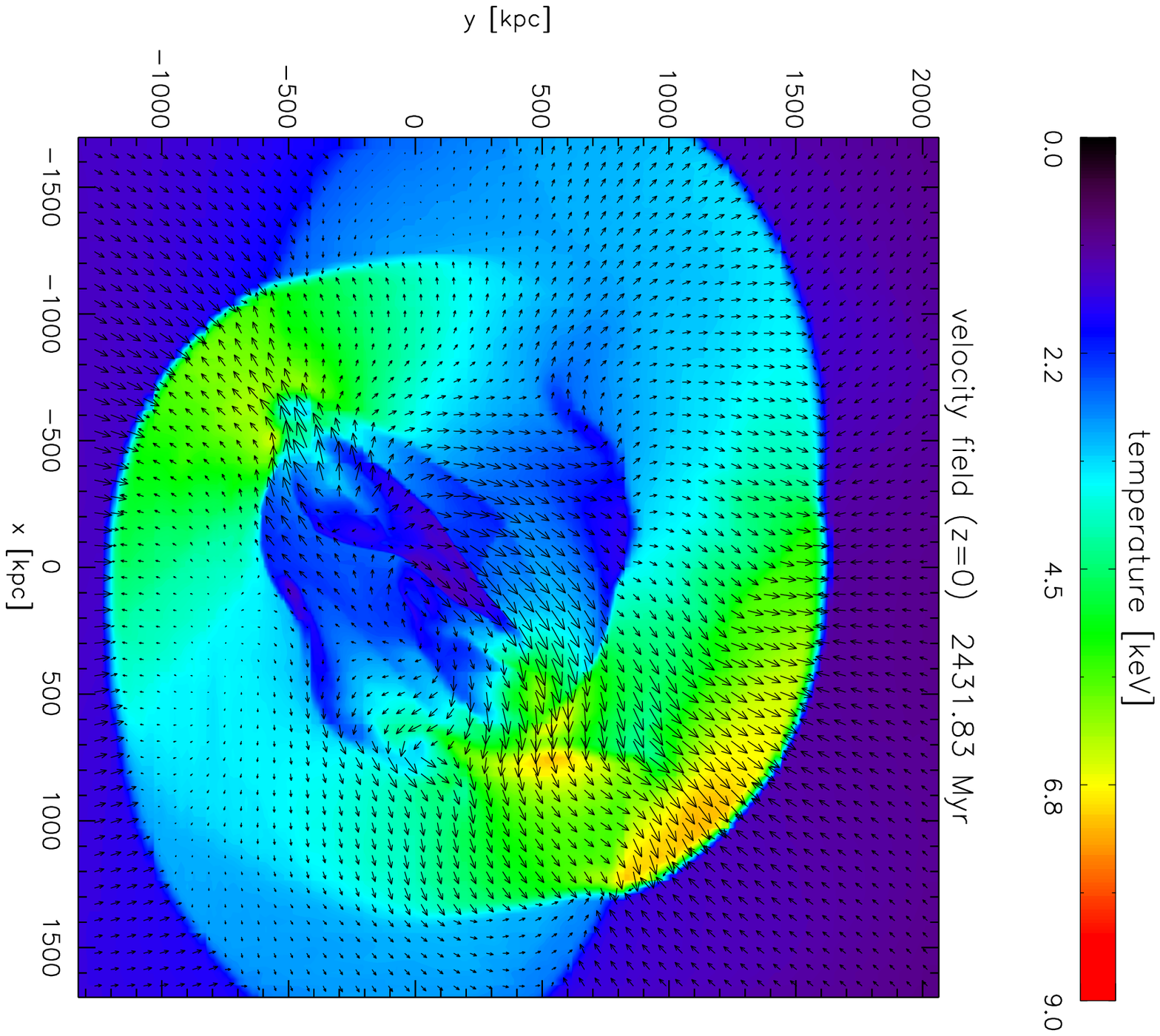}
\caption{Velocity field for a 2:1 mass ratio merger with an impact parameter $b=0$ (left) and $b=536$~kpc (right), for the slice $z=0$. The two snapshots were chosen such that the center of the two shock waves are separated by 2.8~Mpc. The color image displays the temperature distribution. The maximum absolute velocity is $\sim1400$~km~s$^{-1}$.} 
\label{fig:vfield}
\end{figure*}
We used the FLASH 3.2 framework \citep{2000ApJS..131..273F} to simulate collisions between two galaxy clusters, see Fig.~\ref{fig:vfield}. We included standard hydrodynamics and gravity. We simulated a box with a size of $5 \times 5 \times 5$~Mpc and start with two spherical symmetric subclusters with masses $M_1$ and $M_2$ separated by a distance $d$. 
For the gravitational potential of each subcluster we assume hydrostatic equilibrium and spherical symmetry, and we move the center of the gravitational potential of the merging subcluster around the fixed potential of the main cluster, ignoring the interactions between the dark matter.

\begin{figure*}
\includegraphics[angle =90, trim =0cm 0cm 0cm 0cm,width=0.49\textwidth, clip=true]{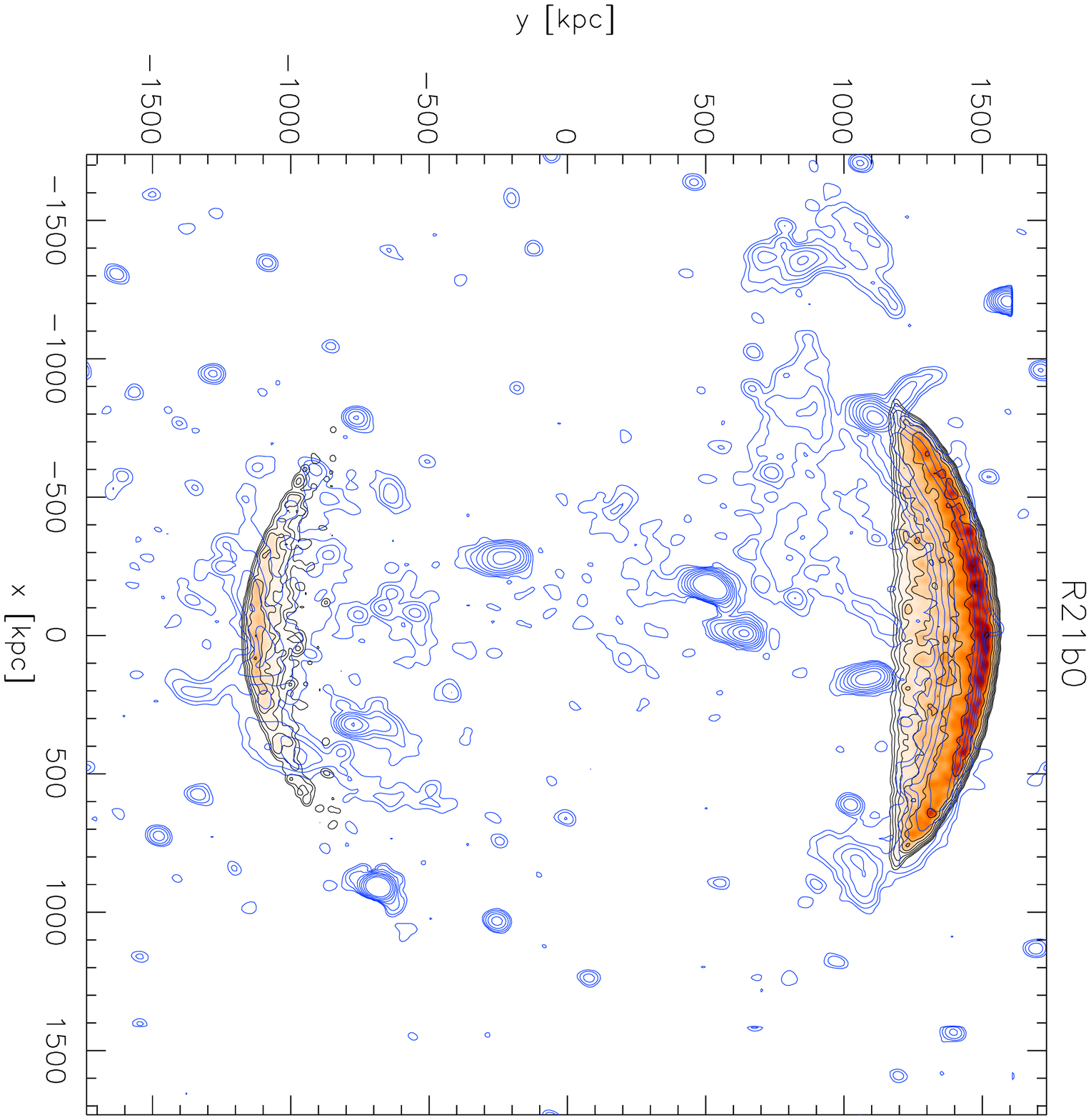}
\includegraphics[angle =90, trim =0cm 0cm 0cm 0cm,width=0.49\textwidth, clip=true]{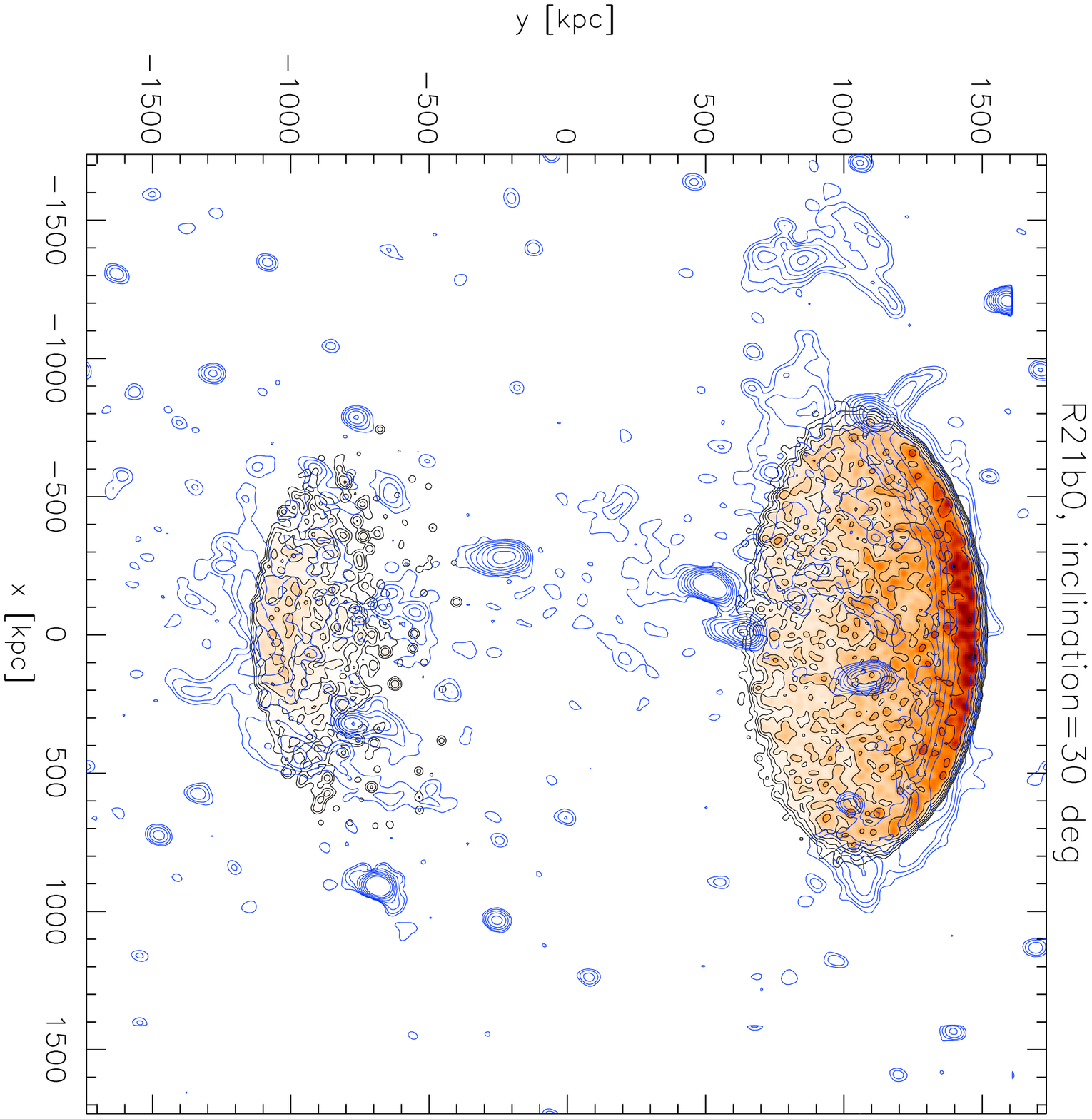}
\caption{Simulated double radio relics. The radio maps  are shown when the two shocks are separated by 2.8~Mpc. Blue contours show the observed emission at 1382~MHz from CIZA~J2242.8+5301 with the WSRT \citep{2010Sci...330..347V}. The orange colored image and black contours display the synthetic radio image. Left: radio map for a $2:1$ mass ratio merger event with zero impact parameter seen edge-on. Right: same as the left panel but the shocks are seen under an angle of $30\deg$ from edge-on.} 
\label{fig:radiomap}
\end{figure*}

Passive tracer particles are used to model the radio emission. At the start of the simulation the particles are distributed according to the density. For particles that pass through a shock we record the Mach number, compression ratio, entropy ratio, and the time since it has passed through the shock. The injection radio spectral index ($\alpha_{\rm{inj}}$) is taken from the Mach number of the shock \citep[e.g.,][]{2008A&A...486..347G}. The radio fluxes are normalized using the method described by \cite{2007MNRAS.375...77H}, which takes into account the efficiency of acceleration as function of Mach number. We adopt constant values for the magnetic field of $B=5.0$~$\mu$G \citep{2010Sci...330..347V}.  With these parameters we then compute the synchrotron emission at a given frequency using the \cite{1973A&A....26..423J} model \citep[see also,][]{1970ranp.book.....P,1994A&A...285...27K}, taking into account the spectral ageing due to synchrotron and inverse Compton losses. A radio map is then simply computed by integrating the radio emission from each tracer particle in the computational volume along a chosen line of sight. Two examples of these radio maps are shown in Fig.~\ref{fig:radiomap}.

\section{Summary}
\label{sec:summary}
We have  simulated radio relics  from idealized binary cluster merger events, varying the mass ratios, impact parameters and viewing angles. The radio emission in these simulations is produced by relativistic electrons which are accelerated via the DSA mechanism.

We will use these simulated radio maps to constrain the merger scenario for the cluster CIZA~J2242.8+5301, which hosts a Mpc scale double radio relic. A preliminary analysis indicates that a merger event with a mass ratio of $1.5:1-2.5:1$ provides the best match to the observed radio emission in this cluster at 1382~MHz. The impact parameter of the merger is constrained to be $b \lesssim 400$~kpc, while the shock fronts (and relics) are observed under and angle less than $\sim10\deg$ from edge-on. The relics are seen at a time of about 1~Gyr after core passage. Our simulations suggest that double radio relics provide a powerful tool to determine the merger parameters when X-ray observations are not available. One of the main uncertainties is the structure and strength of magnetic field in the ICM. Magneto-hydrodynamical simulations of cluster mergers will be needed to study the evolution of the magnetic field and its influence on the radio emission from merger shocks.

\bibliography{ref.bib}
\label{lastpage}
\end{document}